\documentstyle[twoside,fleqn,epsfig,espcrc2]{article}

\newcommand{\AmS}{{\protect\the\textfont2
  A\kern-.1667em\lower.5ex\hbox{M}\kern-.125emS}}

\title{A pot of gold at the end of the cosmic ``raynbow''?}

\author{L. A. Anchordoqui\address{Department of Physics, Northeastern  
University, Boston, MA 02115, USA}\thanks{doqui@hepmail.physics.neu.edu}, 
        M. T. Dova\address{Departamento de F\'{\i}sica, UNLP, 
        CC 67 (1900) La Plata, Argentina}\thanks{dova@venus.fisica.unlp.
        edu.ar},
        T. P. McCauley$^a$\thanks{mccauley@hepmail.physics.neu.edu}, 
        T. Paul$^a$\thanks{tom.paul@hepmail.physics.neu.edu},
        S. Reucroft$^a$\thanks{stephen.reucroft@cern.ch},  
        and J. D. Swain$^a$\thanks{john.swain@cern.ch}}

\begin{document}

\begin{abstract}
We critically review the common belief that ultrahigh energy
cosmic rays are protons or atomic nuclei with masses not exceeding
that of iron. We find that heavier nuclei are indeed possible, and
discuss possible sources and acceleration mechanisms for such
primaries. We also show detailed simulations of extensive air
showers produced by ``superheavy'' nuclei, and discuss prospects
for their detection in future experiments.
\end{abstract}

\maketitle


\newpage

The unambiguous detection of cosmic rays (CRs) with energies above
10$^{20}$ eV (see \cite{yoshida-dai} for a survey and bibliography
on the subject) is a fact of outstanding astrophysical interest.
As shown in the pioneering works of Greisen, Zatsepin, and
Kuzmin \cite{gzk}, the possible sources and the accelerating mechanisms
are constrained by the observed particle spectra due to the
interaction with the universal radiation and magnetic fields on
the way to the observer. The low flux of particles at the end of
the spectrum (the typical rate of CRs above 10$^{20}$ eV is one
event/km$^2$/century) puts strong demands on the collection power
of the experiments, such as can only be achieved by extended air
shower detection arrays at ground level. This indirect method of detection
bears a number of serious difficulties in determining the energy,
mass and/or arrival direction of the primary particles. 

Astrophysical mechanisms to accelerate particles
to energies of up to $10^{21-22}$ eV have been identified, 
but they require exceptional sites \cite{bs}.
Very recently, we have presented a comprehensive study of a
possible nearby 
superheavy-nucleus-zevatron \cite{neu3}.  
We have shown that it is likely that nuclei heavier than iron with
energies above a few PeV can escape from the dense core of a nearby
starburst galaxy like M82, and  eventually be re-accelerated to superhigh 
energies ($E \geq 10^{20}$ eV) at the terminal shocks of galactic 
superwinds generated by the starburst.\footnote{It is important to stress
that M82 is positioned close to the arrival 
direction of the highest CR event detected on Earth. This was first 
pointed out in \cite{ES}.} 
This mechanism 
improves as the charge number $Z$ of the particle is increased. 
Furthermore, we have also shown that the nuclei may arrive on Earth.  
Strictly speaking, the energetic nucleus is seen to lose 
energy mainly as a result of its photodisintegration. In the universal 
rest frame (in which the microwave backgroud radiation is at $3K$), 
the disintegration 
rate $R$ of an extremely high energy nucleus with Lorentz factor $\Gamma$, 
propagating through an isotropic soft photon background of density $n$ 
is given by \cite{stecker},
\begin{equation}
R = \frac{1}{2\Gamma^2} \int_0^\infty dE\, \frac{n(E)}{E^2}
\int_0^{2\Gamma E} dE' \, E'\, \sigma(E'),
\end{equation}
where primed quantities refer to the rest frame of the nucleus, and  
$\sigma$ stands for the total photon absortion cross section. Above 
$10^{20}$ eV, the energy losses are dominated by collisions 
with the relic photons. The fractional energy loss 
around this energy is $R \sim 10^{-15}$ s$^{-1}$. With this in mind, it 
is straightforward to check that superheavy nuclei may impact on 
Earth (for details see Fig. 2 of Ref. \cite{neu3}). In the rest of this report 
we shall discuss the characteristics of the extensive air showers that 
these nuclei may produce after interaction with the atmosphere. 
 
{\it Golden Shower Simulations}:  In order to perform the simulations we 
shall adopt the superposition model.
This model  assumes that an average shower produced by a nucleus with energy 
$E$ and mass number $A$ is indistinguishable from  a superposition of $A$ 
proton showers, each with energy $E/A$. We have generated several sets 
of $^{197}$Au air 
shower simulations by means of the {\sc aires} Monte Carlo 
code \cite{sergio}.\footnote{It should be stressed that for $A> 140$ 
the bulk solar--system abundance 
distribution peaks at $A=195$ \cite{anders-grevesse}. To make some 
estimates, we then  refer our calculations to a gold nucleus.} The 
sample was distributed in the energy range  of $10^{18}$ up to $10^{20.5}$ eV.
{\sc sibyll} was used to  reproduce hadronic collisions above 
200 GeV \cite{sibyll}. All shower particles with
energies above the following thresholds were
tracked: 750 keV for gammas, 900 keV for electrons and positrons, 10
MeV for muons, 60 MeV for mesons and 120 MeV for nucleons and nuclei.
The particles were injected vertically at the top of the atmosphere (100
km.a.s.l), and the surface detector array was put at a depth of  
1036 g/cm$^2$, 
i.e., at sea level. Secondary particles of different types and all charged
particles in individual showers were sorted according to their distance $R$
from the shower axis.

\begin{figure}[htb]
\vspace{9pt}
\epsfig{file=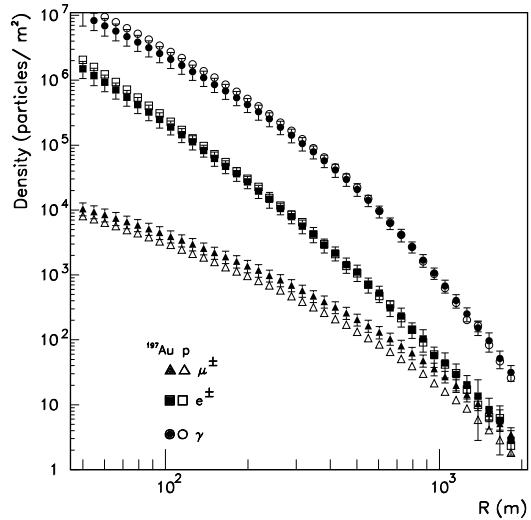,width=7.cm,clip=}
\caption{Ground lateral distributions of 
proton and $^{197}$Au air-showers. The incident energy is 
$E = 3 \times 10^{20}$ eV.}
\label{fig1}
\end{figure}

In Fig. 1 we show the lateral distributions of different groups of secondary 
particles (we have considered separately $\gamma$, $e^+e^-$, 
and $\mu^+\mu^-$).  One can see that the number of 
muons from the gold nucleus shower is greater than the number of muons from 
the proton shower. 

As the cascade develops in the atmosphere, it grows until a maximum size
(number of particles) is reached. The location in the atmosphere where
the cascade has developed the maximum size is denoted by $X_{\rm max}$, with 
units of g cm$^{-2}$. For cascades of a given total energy, heavier 
nuclei have smaller $X_{\rm max}$ than nucleons because the shower is 
already subdivided into $A$ nucleons when it enters the atmosphere. 
At $10^{20}$ eV, the $<X_{\rm max}>$ of a proton (gold) shower is 
$\approx 879$ g/cm$^2$ ($\approx 777$ g/cm$^2$). 
A dust-grain has an even larger cross section, so it tends 
to interact sooner than protons and nuclei \cite{neu1}. In Fig. 2, we 
compare the longitudinal profile of showers initiated by a proton, 
a gold-nucleus and a dust-grain. It is clearly seen how the $X_{\rm max}$  
decreases when increasing the mass. The simulated gold shower is partially 
consistent with the Fly's Eye data. Furthermore, its longitudinal development 
better reproduces the data than protons or dust-grains. It should be remarked,
however,  that extensive air shower simulation depends 
on the hadronic interaction model \cite{hi-prd}. We also point out that for 
the simulation detector effects were not taken into account.

\begin{figure}[htb]
\epsfig{file=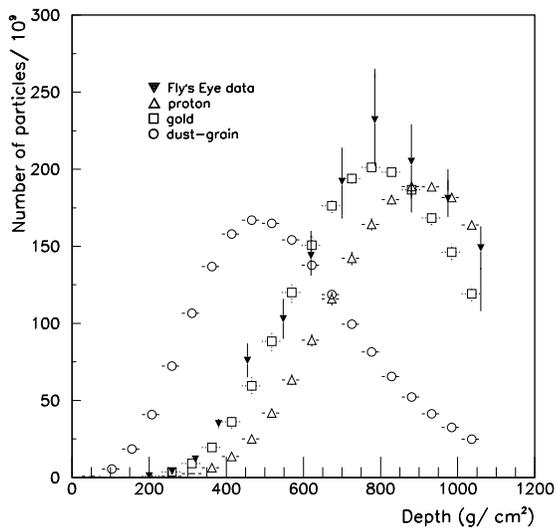,width=7.cm,clip=}
\caption{Longitudinal development of $3\times 10^{20}$ eV showers induced by  
a proton, a gold-nucleus and a dust-grain (log$\,\Gamma = 4.5$), together 
with the data of the highest event recorded by Fly's Eye \cite{FE}.}
\label{fig2}
\end{figure}

Even though the superheavy nucleus hypothesis is partially supported by 
data from the CASAMIA experiment \cite{casamia}, more data is certainly 
needed to verify this model.  In order to significantly increase the 
statistics at the end 
of the spectrum, the Southern Auger Observatory is  currently under 
construction \cite{auger}. It will consist
of a  surface array which will record the lateral and temporal distribution 
of shower particles, and an optical air fluorescence detector, which will 
observe the air shower development in the atmosphere. These two 
techniques provide complementary methods of extracting the required 
information from the shower to test the ideas discussed in this paper.  

\vspace{1cm}

This work was partially supported by CONICET, Fundaci\'on Antorchas and the
National Science Foundation.

\end{document}